\documentclass[conference]{IEEEtran}

\usepackage{array, calc, color, multicol}
\usepackage{algorithm}
\usepackage[noend]{algorithmic} 
\usepackage{graphics, epstopdf, graphicx, epsfig, psfrag, epsf, subfigure}
\usepackage{amssymb, amsfonts, amsmath, times}
\usepackage{relsize}
\usepackage{multicol,balance, verbatim}
\usepackage{textcomp, mathrsfs, stmaryrd, setspace}
\usepackage{amssymb}

\newcommand{\ie}{{\em i.e., }}
\newcommand{\Ie}{{\em I.e., }}
\newcommand{\eg}{{\em e.g., }}
\newcommand{\Eg}{{\em E.g., }}

\newcommand{\modelI}{{\tt{Model I}}}
\newcommand{\modelII}{{\tt{Model II}}}

\newtheorem{example}{Example}

\DeclareGraphicsExtensions{.eps, .pdf, .png, .jpg}   

\IEEEoverridecommandlockouts

\begin{document}

\title{Blocking Avoidance in Transportation Systems}

\author{Shanyu Zhou\\
{\small University of Illinois at Chicago}\\
{ \small \tt szhou45@uic.edu}\\
\and
Hulya Seferoglu\\
{\small University of Illinois at Chicago}\\
{ \small \tt hulya@uic.edu}\\
}

\maketitle

\IEEEpeerreviewmaketitle

{$\hphantom{a}$}\vspace{-30pt}{}

\allowdisplaybreaks

\begin{abstract}

The blocking problem naturally arises in transportation systems as multiple vehicles with different itineraries share available resources. In this paper, we investigate the impact of the blocking problem to the waiting time at the intersections of transportation systems. We assume that different vehicles, depending on their Internet connection capabilities, may communicate their intentions (\eg whether they will turn left or right or continue straight) to intersections (specifically to devices attached to traffic lights). We consider that information collected by these devices are transmitted to and processed in a cloud-based traffic control system. Thus, a cloud-based system, based on the intention information, can calculate average waiting times at intersections. We consider this problem as a queuing model, and we characterize average waiting times by taking into account (i) blocking probability, and (ii) vehicles' ability to communicate their intentions. Then, by using average waiting times at intersection, we develop a shortest delay algorithm that calculates the routes with shortest delays between two points in a transportation network. Our simulation results confirm our analysis, and demonstrate that our shortest delay algorithm significantly improves over baselines that are unaware of the blocking problem.
\end{abstract} 

\vspace{-15pt}
\section{\label{sec:intro} introduction}
In today's metropolitan transportation systems, congestion is one of the major problems. Traffic congestion causes delayed travel times as well as more energy consumption. In fact, the average of yearly delay per auto commuter due to congestion was 38 hours, and it was as high as 60 hours in large metropolitan areas in 2011 \cite{urban_mobility_report}. The congestion caused 2.9 billion gallons of wasted fuel in 2011, and this figure keeps increasing yearly \cite{urban_mobility_report}, \eg the increase was 3.8\% in Illinois between years 2011 and 2012, \cite{illinois_mobility}. This trend poses a challenge for efficient transportation systems in terms of delay and energy, and new congestion control mechanisms are needed to eliminate this inefficiency.

A straightforward approach to address the congestion problem is to enhance the capacity of transportation systems, which requires significant investment. On the other hand, it is extremely important to understand the capacity of existing as well as future transportation systems so that (i) available resources are effectively and fully utilized, and (ii) new transportation systems are developed based on the actual need. Capacity characterization of transportation systems and utilizing available capacity are getting increasing interest recently \cite{dresner}, \cite{papageorgiou}, \cite{gregoire}. This is thanks to vehicle automation, which makes utilization of available capacity possible with the communication and coordination abilities of vehicles, enabled by the Internet connection \cite{dresner}, \cite{papageorgiou}, \cite{gregoire}. However, in this context, it is crucial to take into account practical constraints that arise from real transportation systems while characterizing capacity to fully utilize underlying resources in transportation systems. One of such practical problems is the blocking problem.

In this paper, we investigate the impact of the blocking problem to transportation systems. The blocking problem naturally arises in transportation systems as multiple vehicles with different itineraries share available resources. For example, there may be multiple vehicles that would like to continue straight at an intersection, and they can block the other vehicles that would like to turn left. The next example illustrates the blocking problem using a canonical example.

\begin{figure} [t!]
\centering
\vspace{-10pt}
\includegraphics[width=2.1in]{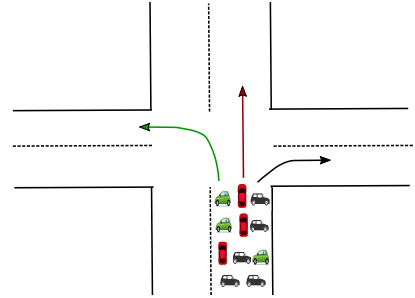}
\caption{An example intersection with multiple vehicles with different routes.}
\vspace{-5pt}
\label{fig:intersection}
\vspace{-15pt}
\end{figure}

\begin{figure}[t!]
\begin{center}
\vspace{-10pt}
\subfigure[Single-lane queue]{ \scalebox{.65}{\includegraphics{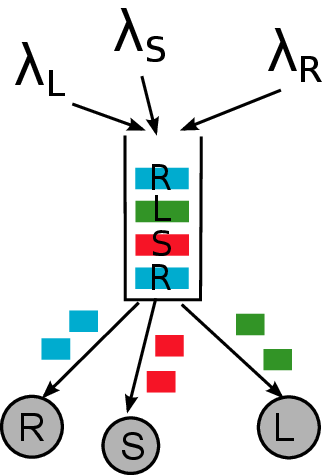}} } \hspace{15pt}
\subfigure[One+Two-lane queue]{ \scalebox{.65}{\includegraphics{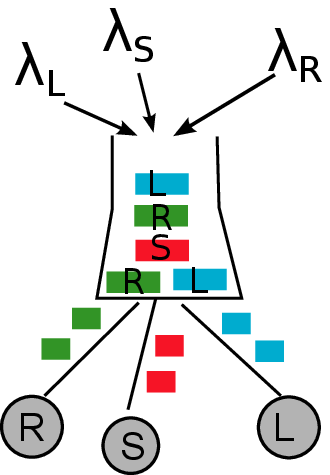}} } \hspace{15pt}
\subfigure[Two+Three-lane queue]{ \scalebox{.65}{\includegraphics{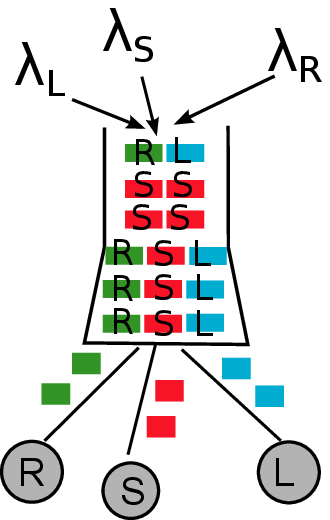}} }
\end{center}
\begin{center}
\vspace{-10pt}
\caption{\label{fig:queueModels} Representation of the south-north bound queue in the intersection demonstrated in Fig.~\ref{fig:intersection} using (a) single-lane, (b) one+two-lane, and (c) two+three-lane queuing models. $\lambda_L$, $\lambda_S$, and $\lambda_R$ are the arrival rates of vehicles to the queue with destinations on the left, straight, and right, respectively.}
\vspace{-25pt}
\end{center}
\end{figure}

\begin{example} \label{example:ex1}
Let us consider Fig. \ref{fig:intersection}, which is an example intersection, where vehicles may turn left, go straight, or turn right.
Fig. \ref{fig:queueModels}(a) is a simple queue representation of Fig. \ref{fig:intersection}, where one-way single-lane traffic is represented as a first-in-first-out (FIFO) queuing mechanism, and head-of-line (HoL) vehicle corresponds to the first vehicle in the queue. Let us assume that the HoL vehicle would like to turn right, but left-turn phase is on. In this case, the HoL vehicle blocks the other vehicles that are waiting in the queue, \ie at the intersection. Similar blocking behavior could also be observed in more realistic scenarios such as in Fig. \ref{fig:queueModels}(b), where lanes are dedicated to turn right (and go straight) or turn left. For example, in  Fig. \ref{fig:queueModels}(b), let us assume that two vehicles are allowed to pass through an intersection when light is on. However, HoL vehicles would like to turn left and right, and the vehicle just behind the HoL vehicles would like to go straight. In this case, if left or right light is on, one vehicle can pass, if go-straight light is on, no vehicle can pass. Thus, even though two vehicles are allowed to pass, only one vehicle can pass in the best-case scenario. A similar argument holds for the example in Fig. \ref{fig:queueModels}(c). As can be seen, blocking occurs in transportation systems, and it increases waiting time and wastes energy. \hfill $\Box$
\end{example}

In this paper, we analyze the effect of the blocking problem to waiting time (which is related to energy consumption) at the intersections of transportation systems. We assume that different vehicles, depending on their Internet connection capabilities, may communicate their intentions (\eg whether they will turn left or right or continue straight) to intersections (specifically to devices attached to traffic lights). We consider that information collected by these devices are transmitted to and processed in a cloud-based traffic control system. Thus, a cloud-based system, based on the intention information, can calculate waiting times at intersections. This calculation would be deterministic if all vehicles could communicate their intentions. However, this is not possible today as only a percentage of devices may have such communication infrastructure. Even in a futuristic setup, where all devices have ability of communicating their intentions, some people may prefer not to share this information due to privacy concerns.  Thus, we consider a hybrid system model, where arbitrary number of devices can communicate their intentions. In this setup, our goal is to (i) calculate average waiting time by taking into account the blocking problem, and (ii) find shortest routes/paths in terms of waiting times. The following are the key contributions of this work:
\begin{itemize}
\item We investigate the impact of blocking problem in transportation systems by modeling arriving and departing vehicles at an intersection as a queuing model. In particular, we investigate two queuing models; single-lane model in Fig. \ref{fig:queueModels}(a) and one+two lane model in Fig.~\ref{fig:queueModels}(b). For each model, we characterize average waiting times by taking into account the vehicles that can communicate their intentions (to turn left, right, or go straight) and blocking probability.
\item We design an algorithm that finds the routes (or set of intersections) between a starting and ending points with shortest delay. The shortest delay algorithm that we design takes into account the average waiting times at intersections, hence blocking probabilities.
\item We evaluate our algorithm via simulations for a multiple-intersection transportation network. The simulation results confirm our analysis, and show that our shortest delay algorithm significantly improves over blocking-unaware schemes.
\end{itemize}

The structure of the rest of this paper is as follows.  Section \ref{sec:related} presents the related work. Section \ref{sec:system model} presents the system model. Sections \ref{sec:waiting_time} characterizes average waiting time at the intersections by taking into account the blocking problem. Sections \ref{sec:algorithm} develops a shortest delay algorithm. Section \ref{sec:sims} presents simulation results. Section \ref{sec:conclusion} concludes the paper.

\vspace{-5pt}
\section{related work}\label{sec:related}
Analyzing waiting times and modeling transportation systems using queueing theory have a long history (more than 50 years) \cite{webster}. \Eg \cite{miller}, \cite{newell}, \cite{dirk} considered one-lane queues and calculated the expected queue length and arrivals using probability generation functions. These models focus on fixed-cycle traffic signals, and they calculate the steady-state delays and queue lengths under the assumption that the arriving process does not change over time \cite{pitu_ning}. Time-dependent arrivals have also been considered in \cite{yang_xianfeng}, \cite{kimber}, \cite{akcelik}, \cite{akcelik_rouphail}, \cite{hunt}, \cite{lowrie}, \cite{henry}, \cite{mirchandani}, \cite{gartner}, \cite{diakaki}. Different modeling strategies are also studied; such as the queuing network model \cite{osorio}, cell transmission model \cite{lo}, store-and-forward \cite{aboudolas}, and petri-nets \cite{febbraro}. As compared to this line of work, our work considers  (i) that some vehicles can communicate their intentions, which affects the delay analysis, and (ii) blocking problem.

Recently, with the development of sensor technology, there is an increasing interest in terms of estimating the average delay and queue length at isolated intersections using the information collected by probing vehicles. The queue length estimation and estimation error are characterized in \cite{gurcan_mecit}, \cite{mandoye_virgil}, where the probing vehicles can provide their location and entering and departure time in the road intersection to a central controller. Isolated intersection is considered in  \cite{pitu_ning}, and optimal traffic cycle to adaptively serve two directions of vehicles are calculated. An optimization scheme is proposed in \cite{guo_yin} by taking into account stochastic features of traffic flows in isolated intersections. As compared to this line of work, we focus on the impact of blocking problem to waiting times in transportation systems.

There is also an increasing interest in terms of controlling transportation networks by analyzing the network as a whole without focusing on specific intersections. For example, the control of a network of signalized intersections is discussed in \cite{gregoire}, with the goal of stabilizing waiting times at intersections by considering the network as a whole and directing vehicles to appropriate intersections. However, this line of work does not take into account the blocking probability and its impact to the waiting times and queue sizes.

\vspace{-5pt}
\section{system model}\label{sec:system model}
We consider a transportation network consisting of multiple intersections. First, we will focus on an intersection as exemplified in Fig.~\ref{fig:intersection} as isolated from the rest of the network, and characterize waiting times by taking into account blocking problem. Then, we will consider multiple intersections in a network, and determine the shortest delay routes/paths.

In our setup, we model each intersection as a set of queues. For example, in Fig.~\ref{fig:intersection}, there are four queues for each direction (for south-north, north-south, west-east, and east-west bounds). We specifically focus one direction in our model; \eg south-north bound in Fig.~\ref{fig:intersection}, and model it using two models: \modelI, which is one-lane model shown in Fig.~\ref{fig:queueModels}(a) and \modelII; which is a one+two lane model shown in Fig.~\ref{fig:queueModels}(b).

We consider a time-slotted system, where one vehicle is allowed to leave the queue during one time slot. At each slot, the arrivals of vehicles are distributed according to Poisson distribution. In this setup, when a vehicle enters a queue, it can communicate with the intersection and inform its intention in terms of turning left or right or continuing straight. We call this ``communication ability'' of the vehicle and, we call the probability that a vehicle has communication ability is ``communication probability''. \Ie each vehicle can communicate with  the intersection (actually with the sensor possibly attached to traffic lights) with some ``communication probability''. In this setup, if vehicles at an intersection have communication abilities, then the traffic light can arrange its phases accordingly, so {\em blocking is avoided}. Otherwise, the traffic light selects a phase randomly, and {\em blocking may occur}.

Now, let us explain our queuing models; \modelI~ and \modelII~ in detail.

\vspace{-5pt}
\subsection{\modelI}
\modelI~ is a queuing model illustrated in Fig.~\ref{fig:queueModels}(a). In this setup, $\lambda_L$, $\lambda_S$, and $\lambda_R$ correspond to arrival rates of vehicles that would like to turn left, continue straight, and turn right. However, for simplicity, we will assume in the rest of the paper that $\lambda_1 = \lambda_L$, $\lambda_2 = \lambda_S$, and $\lambda_3 = \lambda_R $.

In this setup, if the head-of-line (HoL) vehicle at time slot $t$ has communication ability, then there is no blocking in slot $t$. However, if the HoL vehicle does not have communication ability, traffic phases for left, straight, and right are ``ON'' with probability $p_1$, $p_2$, and $p_3$, respectively. Thus, if HoL vehicle would like to turn left, but traffic phase turn right is ``ON'', then blocking occurs. We consider in our model that $p_i$ ($i=1,2,3$) is pre-determined and independent and identically distributed (i.i.d) over time slots.

In this setup, arrival traffic is Poisson and departure is any general traffic. Thus, we can model this queue according to M/G/1 queue. However, the analysis is not straightforward due to blocking effect. We present the details in Section~\ref{sec:waiting_time}.

\vspace{-5pt}
\subsection{\modelII}
Our second queuing model is \modelII, which is shown in Fig.~\ref{fig:queueModels}(b). In this setup, to demonstrate the analysis in a simple way, we consider that the right-turning and continuing-straight traffics are combined together. Thus, we consider the arrival rates as $\lambda_1 = \lambda_R + \lambda_S$ and $\lambda_2 = \lambda_L$.

\begin{figure} [t!]
\vspace{-10pt}
\centering
\scalebox{.77}{\includegraphics{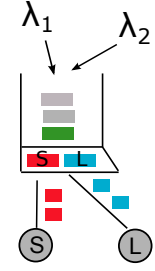}}
\caption{\label{model2} The queuing model with two vehicles in the head-of-line (HoL). The arrival rates of right-turning and straight-continuing vehicles are merged as $\lambda_1$ and the arrival rate of left-turning vehicle is $\lambda_2$.}
\vspace{-15pt}
\end{figure}

In this model, there are two dedicated lanes for vehicles at the head-of-line (HoL), where left-turning vehicles can only join the left lane, while the right-turning or straight-continuing vehicles can only join the going straight lane. Note that to further simplify the terminology, we call both right-turning and straight-continuing vehicles as going-straight vehicles in the rest of the paper. The simplified model is shown in  Fig.~\ref{model2}.

In this setup, at any phase, at least one vehicle is transmitted. However, when we consider traffic phases lasting for two slots, either one (if blocking occurs) or two (if there is no blocking) vehicles are transmitted. In this model, we consider the case that traffic phases change at every two slots, and provide analysis for this setup. Similar to \modelI, our goal is to characterize the average waiting time of this model using M/G/1 queues. The details are provided in the next section.

\vspace{-5pt}
\section{Average Waiting Time Analysis}\label{sec:waiting_time}
In this section, we characterize the average waiting time for \modelI~ and \modelII. Both queuing models can be modeled as M/G/1 queues. Thus, the average waiting times should follow the Pollaczek-Khinchine (PK) formula:
\begin{equation} \label{eq:PK}
W=\frac{\sum\limits_{i=1}^{3}\lambda_iE(x^2)}{2(1-\rho)}
\end{equation} where $W$ is the average waiting time, $x$ is the service time and $\rho$ is the line utilization, $\rho=\sum\limits_{i=1}^{3}\lambda_iE(x)$. However, in this formula, it is not straightforward to characterize the expected service time $E(x)$ and the second moment of the service time $E(x^2)$ due to different communication abilities of vehicles and blocking. Thus, our main contribution in this section is the characterization of these parameters ($E(x)$ and $E(x^2)$), which we discuss next.

\vspace{-5pt}
\subsection{Average Waiting Time for \modelI}
In \modelI, if there is a mismatch between the traffic phases and HoL vehicle, then blocking occurs. For example, if HoL vehicle would like to turn left, but traffic phase turn right is ``ON'', then blocking occurs. This kind of blocking occurs only if HoL vehicle does not have communication ability. Thus, we discuss three scenarios with different levels of communication abilities; (i) all vehicles have communication abilities, (ii) none of the vehicles have communication abilities, (iii) a percentage of vehicles have communication abilities. Note that the first two scenarios are actually the special cases of the third scenario. However, we present the average waiting time analysis for these scenarios as well in the following to better explain our approach.

\subsubsection{All vehicles have communication abilities}\label{everycartellsmodel1}
Since every vehicle has communication ability, the traffic light always knows where the HoL vehicle would like to go. Thus, it can arrange the traffic phase accordingly, so the service time $x$ in (\ref{eq:PK}) will be 1. Thus, $E(x)=1$ and $E(x^2)=1$, and (\ref{eq:PK}) is expressed as

\begin{equation} \label{waitingtime_servicerate1}
W=\frac{\sum\limits_{i=1}^{3}\lambda_iE(x^2)}{2(1-\rho)}=\frac{\sum\limits_{i=1}^{3}\lambda_i}{2(1-\sum_{i=1}^{3}\lambda_i)}
\end{equation} where the total arrival rate to the queue is $\sum_{i=1}^{3}\lambda_i$. Note that this is a trivial scenario without any blocking. Next, we consider the case that none of the vehicles have communication abilities.

\subsubsection{None of the vehicles have communication abilities}\label{nocarscantellmodel1}
If none of the vehicles have communication abilities, then the traffic lights randomly choose phases for left, straight or right with probabilities; $p_1$, $p_2$, $p_3$, respectively.

In this setup, the HoL vehicle can take more than one slots to pass the intersection. Therefore, the service time $x$ can be any positive integer. If the service time $x=n$, $\forall n\in \mathbb{R^+}$, this means that there has been a mismatch between the HoL vehicle and the traffic phases in the last $n-1$ slots. Therefore, the probability that the vehicle pass the intersection at the $n$th slot is $P[x=n]$, and it follows the geometric distribution
\begin{equation} \label{eq:eq1}
P[x=n]=\sum\limits_{i=1}^{3}\alpha_i (1-p_i)^{n-1}p_i
\end{equation}
where $\alpha_i$, $(i=1,2,3)$ is the probability that the HoL vehicle is going to left, straight or right. It is straightforward to see that $\alpha_i={\lambda_i}/{\sum\limits_{i=1}^{3}\lambda_i}$. Thus, (\ref{eq:eq1}) is expressed as
\begin{equation} \label{eq:eq2}
P[x=n]=\frac{\sum_{i=1}^{3}\lambda_i(1-p_i)^{n-1}p_i}{\sum_{i=1}^{3}\lambda_i}
\end{equation}
Using (\ref{eq:eq2}), we can calculate $E(x)$ and $E(x^2)$ (the detailed calculations are provided in \cite{this_tech}), which leads to the average waiting time
\begin{equation} \label{waitingtime_model1_noonetells}
W=\frac{\sum\limits_{i=1}^{3}\lambda_iE(x^2)}{2(1-\rho)}=\frac{\sum\limits_{i=1}^{3}\lambda_i\frac{2-p_i}{p_i^2}}{2(1-\sum_{i=1}^{3}\frac{\lambda_i}{p_i})},
\end{equation}

Note that the average waiting time for this scenario directly depends on the traffic phase probabilities. This is intuitive as the mismatch between vehicles and traffic lights increases the waiting time, which directly affects the average waiting time.

\subsubsection{A percentage of vehicles has communication abilities}\label{somecarcantellmodel1}
Now, we assume that a percentage of vehicles has communication abilities. Let $p_t$ denotes the communication probability of a vehicle. In this scenario, when the HoL vehicle has communication ability, then the traffic phases could be arranged accordingly, and the vehicle immediately passes the intersection (\ie it can take 1 slot to pass the intersection). On the other hand, when the vehicle does not have communication ability, then the traffic phases will be left, straight or right randomly with probabilities $p_1$, $p_2$ and $p_3$, respectively.

In this scenario, the service time $x=1$ occurs in two cases. The first case is that the HoL vehicle has communication ability. The second case is that the HoL vehicle does not have communication ability, but the traffic phase is aligned with the direction of the vehicle in the first slot (note that it may take longer). The probability of the second case is $(1-p_t)\sum\limits_{i=1}^{3}\alpha_ip_i=(1-p_t)\frac{\sum_{i=1}^{3}\lambda_ip_i}{\sum_{i=1}^{3}\lambda_i}$. Thus, we can calculate the probability that the service time is equal to  1 as
\begin{equation}
P[x=1]=p_t+(1-p_t)\frac{\sum_{i=1}^{3}\lambda_ip_i}{\sum_{i=1}^{3}\lambda_i}
\end{equation}

If the service time is larger than 1, we know that this only happens when the HoL vehicle does not have communication abilities. Thus, the calculation for $x>1$ case is similar to the scenario that none of the vehicles have communication abilities. Thus, the probability of service time to be $n (n \geq 2)$ is
\begin{equation} \label{eq:pxn_2_2}
P[x=n]=(1-p_t)\frac{\sum_{i=1}^{3}\lambda_i(1-p_i)^{n-1}p_i}{\sum_{i=1}^{3}\lambda_i}
\end{equation} 
Using $P[x=n]$ in (\ref{eq:pxn_2_2}), we can calculate $E(x)$ and $E(x^2)$ as details are provided in \cite{this_tech}. Then the average waiting time is characterized as 
\begin{equation} \label{waitingtime_model1}
W=\frac{\sum_{i=1}^{3}\left[\lambda_ip_t+(1-p_t)\frac{\lambda_i(2-p_i)}{p_i^2}\right]}{2\left\lbrace 1-\sum_{i=1}^{3}\left[\lambda_i p_t+(1-p_t)\frac{\lambda_i}{p_i}\right]\right\rbrace}
\end{equation}


\vspace{-5pt}
\subsection{Average Waiting Time for \modelII}
Now, we consider the characterization of the average waiting time for \modelII~ shown in Fig.~\ref{model2}. In this model, we consider that traffic phases change at every two slots. Thus, in this two-slot duration, one vehicle passes if there is blocking, and two vehicles can pass if there is no blocking. Similar to \modelI, we consider different levels of communication abilities for vehicles; (i) all vehicles have communication abilities, (ii) none of the vehicles have communication abilities, and (iii) a percentage of vehicles has communication abilities. Next, we present our analysis for each scenario.

\subsubsection{All vehicles have communication abilities}\label{everycartellsmodel2}
If every vehicle has communication abilities, two vehicles can pass at every two-slot duration. Thus, the service time for vehicles is always 1. Therefore, the average waiting time will be the same as in (\ref{waitingtime_servicerate1}).

\subsubsection{None of the vehicles have communication abilities}\label{nocarscantellmodel2}
\begin{figure}[t!]
\vspace{-10pt}
\begin{center}
\subfigure[Configuration I]{ \scalebox{.5}{\includegraphics{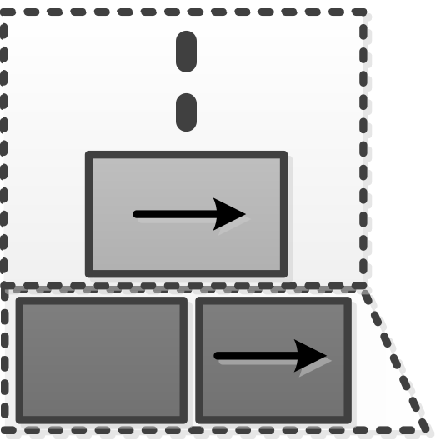}} } \hspace{15pt}
\subfigure[Configuration II]{ \scalebox{.5}{\includegraphics{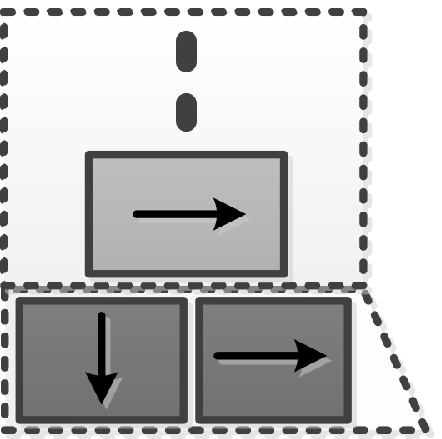}} } \\
\subfigure[Configuration III]{ \scalebox{.5}{\includegraphics{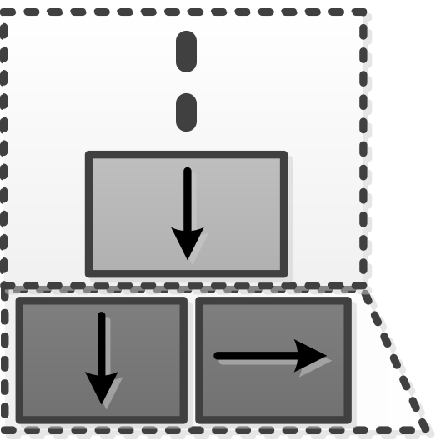}} }  \hspace{15pt}
\subfigure[Configuration IV]{ \scalebox{.5}{\includegraphics{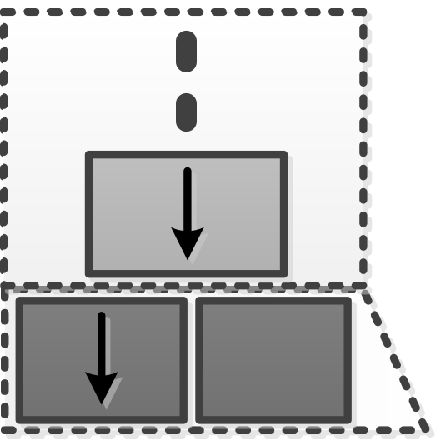}} }
\end{center}
\begin{center}
\caption{\label{fig:holconfiguration} Four possible configurations for the first three vehicles in \modelII. Arrow in each configuration represents the direction of the vehicle while empty space represents that there are no vehicles in that position. (a) Configuration I: HoL locations: {\em empty}, {\em turn-left}. Vehicle behind HoL locations is {\em turn-left} (b) Configuration II: HoL locations: {\em go-straight}, {\em turn-left}. Vehicle behind HoL locations is {\em turn-left}, (c) Configuration III: HoL locations: {\em go-straight}, {\em turn-left}. Vehicle behind HoL locations is {\em go-straight}, (d) Configuration IV: HoL locations: {\em go-straight}, {\em empty}. Vehicle behind HoL locations is {\em go-straight}.}
\vspace{-20pt}
\end{center}
\end{figure}

When none of the vehicles have communication abilities, traffic phases can be arranged depending on the first three vehicle configurations in the queue. Fig. \ref{fig:holconfiguration} shows four possible configurations for these three vehicles. In particular, Fig. \ref{fig:holconfiguration}(a) shows that the configuration for the HoL locations are {\em empty} and {\em turn-left}, and the vehicle just behind the HoL locations is also {\em turn-left}. Note that in this setup, the vehicle behind the HoL locations cannot take a position in the empty HoL position, because this vehicle would like to turn left.

When the first three vehicles in the HoL are in the form of Fig. \ref{fig:holconfiguration}(a) or Fig. \ref{fig:holconfiguration}(d), \ie where one of the HoL positions is empty, we assume that the traffic light can sense whether this location is empty or not. As long as it senses an empty location, \eg an empty left-turning HoL location, it can be estimated that the only possible configuration is Configuration IV in Fig. \ref{fig:holconfiguration}(d). Then, the traffic light can arrange its phase to align with the configuration. On the other hand, if the configurations are Configuration II (Fig. \ref{fig:holconfiguration}(b) ) or Configuration III (Fig. \ref{fig:holconfiguration}(c)), the traffic light will randomly arrange its phases.

\begin{figure}[t!]
\vspace{10pt}
\begin{center}
\subfigure[Conf. II: Left-turn phase ON ]{ \scalebox{.42}{\includegraphics{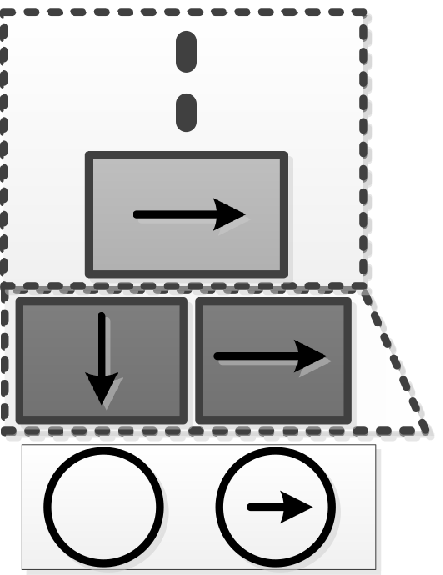}} } \hspace{10pt}
\subfigure[Conf. II: Go-straight phase ON ]{ \scalebox{.42}{\includegraphics{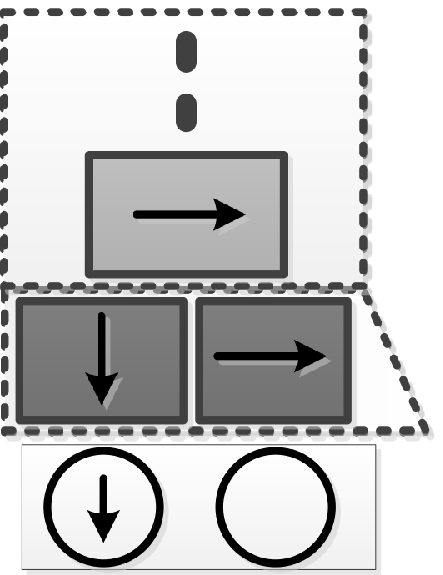}} } \\
\subfigure[Conf. III: Left-turn phase ON ]{ \scalebox{.42}{\includegraphics{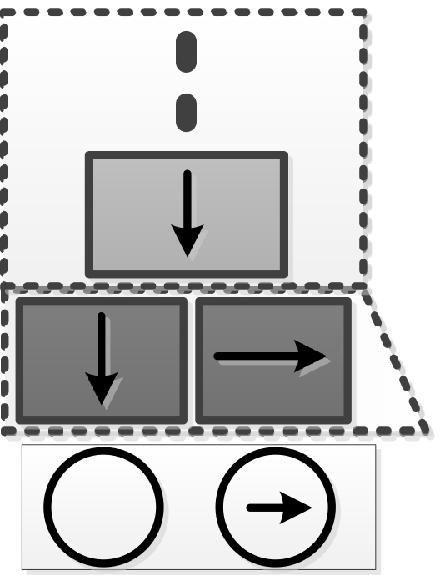}} } \hspace{10pt}
\subfigure[Conf. III: Go-straight phase ON ]{ \scalebox{.42}{\includegraphics{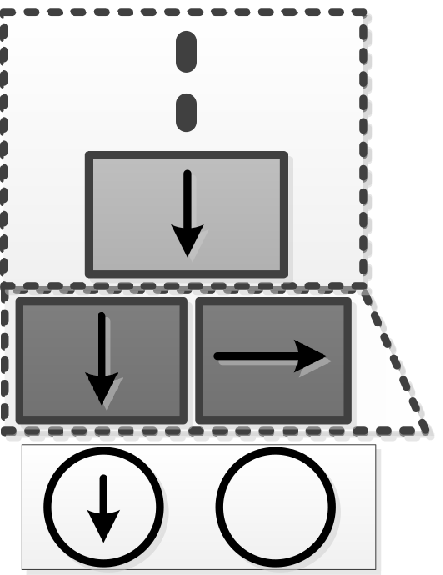}} }
\end{center}
\begin{center}
\vspace{-5pt}
\caption{\label{fig:holconfig_lights} Vehicle configurations expansion; Configuration II and Configuration III in Fig.~\ref{fig:holconfiguration} for different traffic phases. (a) Configuration II: Left turn phase is ON. (b) Configuration II: Go straight phase is ON. (c) Configuration III: Left turn phase is ON. (d) Configuration III: Go straight phase is ON.}
\vspace{-25pt}
\end{center}
\end{figure}

Based on the discussion above, we know that the service time can be 1 when there is an empty HoL location or the traffic phase allows two vehicles go within two slots when there are no empty HoL locations. Since the service time can be 1 or 2 when there are no empty HoL locations, to calculate the expected service time in this case, we can expand Fig. \ref{fig:holconfiguration}(b) and Fig. \ref{fig:holconfiguration}(c) with traffic light status to make the calculation more efficient. The expansion is illustrated in Fig. \ref{fig:holconfig_lights}.

Now, it is clear that the service time will be 1 when the HoL vehicles and traffic light are in the form of Fig. \ref{fig:holconfiguration}(a), Fig. \ref{fig:holconfiguration}(d), Fig \ref{fig:holconfig_lights}(a), or Fig. \ref{fig:holconfig_lights}(d), and will be 2 when they are in the form Fig. \ref{fig:holconfig_lights}(b) or Fig. \ref{fig:holconfig_lights}(c).

\begin{figure} [t!]
\vspace{-15pt}
\centering
\scalebox{.28}{\includegraphics{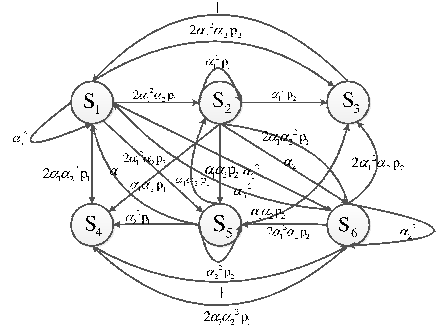}}
\caption{\label{fig:markov} The Markov chain for states $S_i$ ($i=1,\ldots,6$), where $\alpha_i$ ($i=1,2$) denotes the probability that the HoL vehicle is going straight or left and $p_i$ ($i=1,2$) denotes the probability that the traffic light choose go-straight or turn-left phases.}
\vspace{-20pt}
\end{figure}

By taking into account all possible configurations in Fig. \ref{fig:holconfiguration} and Fig. \ref{fig:holconfig_lights}, we can obtain six states of the system. Let $S_1$, $S_2$, $S_3$, $S_4$, $S_5$ and $S_6$ denote these six states. Thus, we can use a Markov chain to calculate the stationary probability of the six states and accordingly, the probability distribution of the service time. The Markov chain is shown in Fig. \ref{fig:markov}.

Now, let $x$ be the random variable representing service time. At stationary state, it is clear that $P[x=2]=P[S_3]+P[S_4]$, where $P[S_i](i=1,\ldots,6)$ is the stationary probability that the system is at state $S_i$. Since the service time can only be 1 or 2, then $P[x=1]=1-P[x=2]$. Solving the balance equations of the Markov chain in Fig. \ref{fig:markov}, we can obtain
\begin{eqnarray}
P[x=2] = \frac{2\lambda_1\lambda_2(\lambda_1p_2+\lambda_2p_1)}{(\lambda_1+\lambda_2)(\lambda_1^2p_2+\lambda_2^2p_1+4\lambda_1\lambda_2)}
\end{eqnarray}
Hence
\begin{eqnarray}
E(x) = 1+\frac{2\lambda_1\lambda_2(\lambda_1p_2+\lambda_2p_1)}{(\lambda_1+\lambda_2)(\lambda_1^2p_2+\lambda_2^2p_1+4\lambda_1\lambda_2)}
\end{eqnarray}
\begin{eqnarray}
E(x^2) =  1+\frac{6\lambda_1\lambda_2(\lambda_1p_2+\lambda_2p_1)}{(\lambda_1+\lambda_2)(\lambda_1^2p_2+\lambda_2^2p_1+4\lambda_1\lambda_2)}
\end{eqnarray}
Therefore, by using PK formula in (\ref{eq:PK}), we can obtain the average waiting time as
\begin{equation} \label{waitingtime_model2_noonetells}
W=\frac{(\lambda_1+\lambda_2)+\frac{6\lambda_1\lambda_2(\lambda_1p_2+\lambda_2p_1)}{\lambda_1^2p_2+\lambda_2^2p_1+4\lambda_1\lambda_2}}{2\left[1-(\lambda_1+\lambda_2)-\frac{2\lambda_1\lambda_2(\lambda_1p_2+\lambda_2p_1)}{\lambda_1^2p_2+\lambda_2^2p_1+4\lambda_1\lambda_2}\right]}
\end{equation}

\subsubsection{A percentage of vehicles has communication abilities}\label{somecarcantellmodel2}
Now, we assume that the probability of a vehicle has a communication ability is $p_t$. A major difference in \modelII~ as compared to \modelI~ is that we have dedicated lanes for left-turning and straight-going vehicles in \modelII. Since we assume that  traffic lights can sense whether a dedicated lane (HoL location) is empty or not, then the first two vehicles in  HoL locations will indirectly communicate their intentions. \ie even if a vehicle does not communicate their intention, when it goes to turn-left lane, then it can be sensed by the traffic light, and its intention can be inferred.

Thus, it only matters whether the third vehicle (the vehicle right behind the HoL locations) has communication ability or not. Note that if the third vehicle has communication ability, then the service time will always be 1 (\ie 2 vehicles can pass in two slots) since traffic phases can be arranged to match the configurations of the first three vehicles. If the third vehicle does not have communication ability, then the probability distribution of service times will be exactly the same as that in section \ref{nocarscantellmodel2}. Therefore, the service time becomes 2 when the third vehicle does not have communication ability and the traffic phase does not match its intention. Similar to the analysis in Section \ref{nocarscantellmodel2} and as detailed in \cite{this_tech}, we have
\begin{equation}\label{waitingtime_model2}
W=\frac{(\lambda_1+\lambda_2)+\frac{6(1-p_t)\lambda_1\lambda_2(\lambda_1p_2+\lambda_2p_1)}{\lambda_1^2p_2+\lambda_2^2p_1+4\lambda_1\lambda_2}}{2\left[1-(\lambda_1+\lambda_2)-\frac{2(1-p_t)\lambda_1\lambda_2(\lambda_1p_2+\lambda_2p_1)}{\lambda_1^2p_2+\lambda_2^2p_1+4\lambda_1\lambda_2}\right]}
\end{equation} 

%

\vspace{-5pt}
\section{Shortest Delay Algorithm}\label{sec:algorithm}
In this section, we consider a transportation system, which consists of multiple intersections and roads. In this setup, by using the average waiting time analysis in Section~\ref{sec:waiting_time}, we develop a shortest delay algorithm.

In particular, we consider the transportation network as a weighted directed graph $G$ with $N$ nodes and $M$ edges, each node represents an intersection and each edge represents a road that connects two intersections. Let $V(G)$, $E(G)$ denotes the set of nodes and edges respectively. Then $|V(G)|=N$, $|E(G)|=M$. The weight of each edge is the sum of the average waiting time in the arrival node (intersection) and the traveling time between the two nodes (intersections) on that edge (road). Given the starting and destination nodes in the graph, the goal of our algorithm is to find a path that returns the shortest waiting time (\ie shortest delay path). The shortest delay algorithm is expressed more specifically in the following.
\begin{itemize}
\item Given the arrival rate of traffic flows into each node $n\in V(G)$ and communication probability of vehicles arriving into that node, we can calculate the average waiting time $W_n$ using the waiting time analysis in Section \ref{sec:waiting_time}. For \modelI, $W_n$ is (\ref{waitingtime_model1}) and for \modelII, $W_n$ is (\ref{waitingtime_model2}).

\item We assume that vehicles travel at a steady speed of $s$ over an edge between two nodes before arriving into the queue/intersection. Thus, the traveling time between two nodes $t_m$ ($m \in E(G)$) can be obtained by dividing the length of the edge $L_m$ by the speed $s$, \ie $t_m=L_m/s$. Then, the total traveling time $T_m$ is the sum of $W_n$ and $t_m$, \ie $T_m=W_n+t_m$, and $T_m$ is assigned as the weight of edge $m$.

\item Finally, given the starting and ending nodes, we run Dijkstra's algorithm and return the shortest path it finds. Since the weight of each edge is the total traveling time on that edge, the shortest path we obtain in this way becomes the shortest delay algorithm.
\end{itemize}

Note that our shortest delay algorithm finds the routes with the shortest average waiting times and less blocking probabilities, so it provides {\em blocking avoidance} over transportation networks. We note that in multiple-intersection transportation networks (with vehicles making decisions based on estimated waiting times), arrival rates of vehicles would be more generic than Poisson arrivals. In this sense, our algorithm in this section is an approximation. In the next section, we show the effectiveness of our algorithm.

\vspace{-5pt}
\section{Performance Evaluation}\label{sec:sims}
In this section, we first consider isolated intersections, and evaluate our waiting time analysis provided in Section~\ref{sec:waiting_time}. Then, we consider a larger transportation network with multiple intersections, and evaluate our shortest delay algorithm developed in Section~\ref{sec:algorithm}.

\vspace{-5pt}
\subsection{Evaluation of Average Waiting Time at Isolated Intersections}

\subsubsection{\modelI}
\begin{figure}[t!]
\vspace{-10pt}
\begin{center}
\subfigure[Waiting time vs arrival rate]{ \scalebox{.30}{\includegraphics{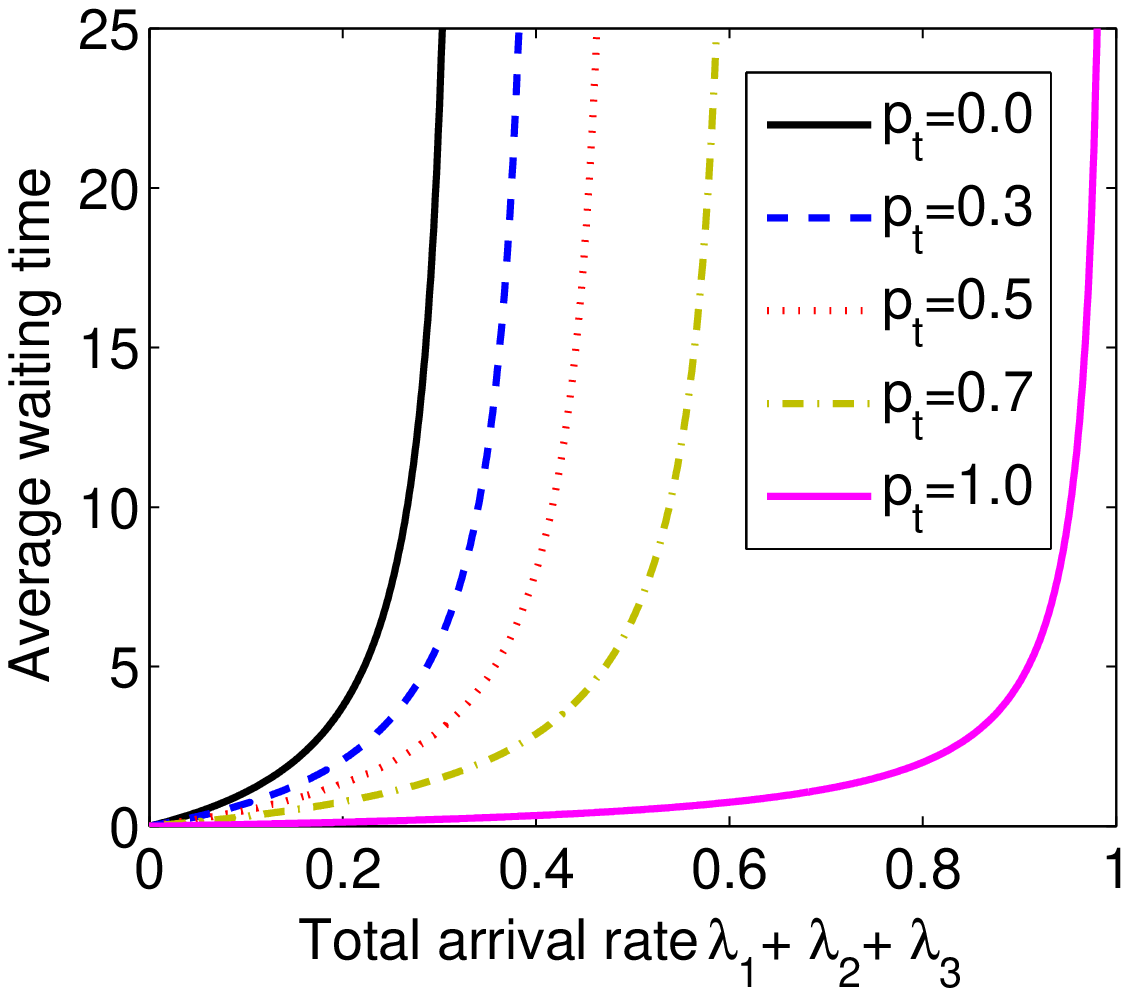}} } 
\subfigure[Queue Size vs $p_t$]{ \scalebox{.30}{\includegraphics{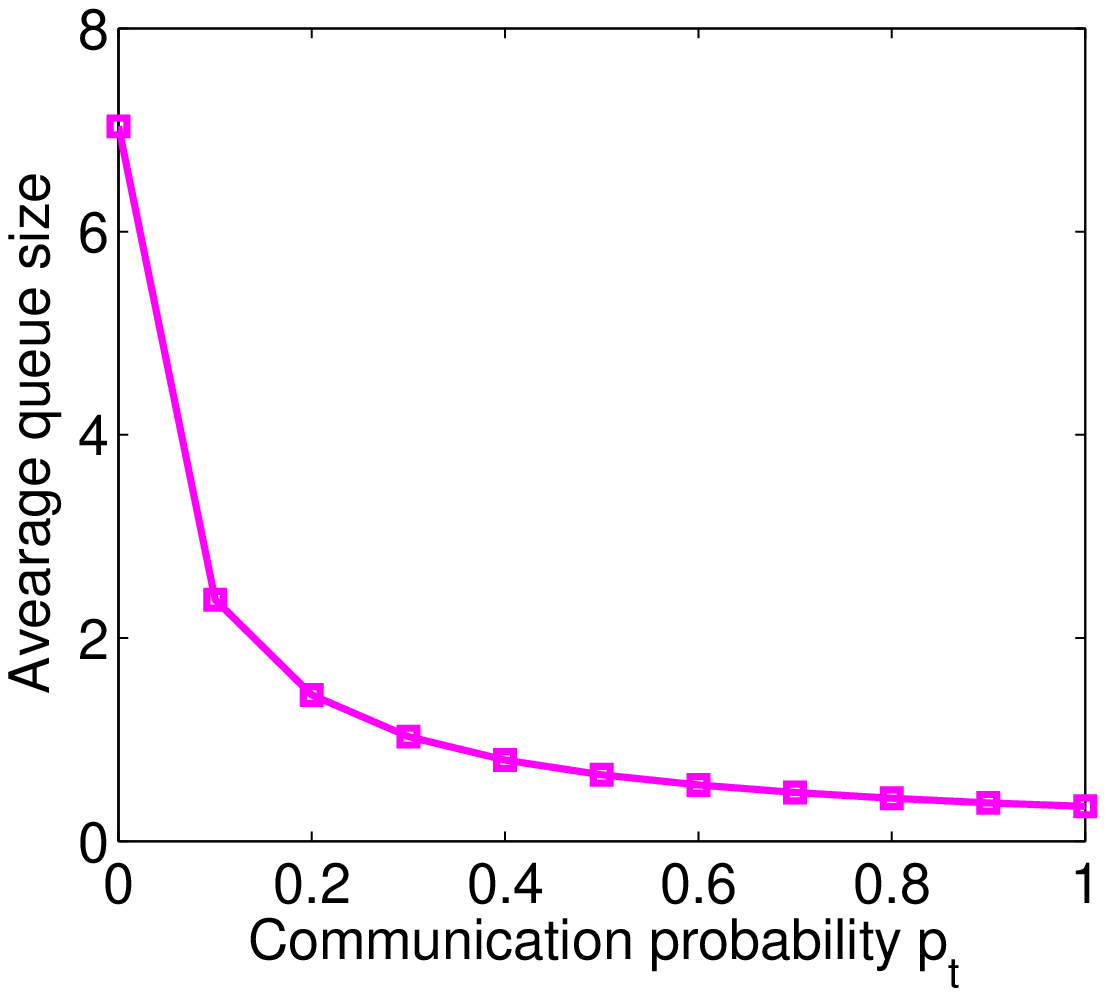}} } \\
\end{center}
\begin{center}
\vspace{-5pt}
\caption{\label{fig:sims_one_queue_modelI} \modelI. (a) Average waiting time for different total arrival rates with different communication probabilities $p_t$. (b) Average queue size versus communication probability $p_t$. The arrival rates are $\lambda_1=\lambda_2=\lambda_3=0.1$.}
\vspace{-25pt}
\end{center}
\end{figure}

The simulated average waiting time for \modelI~ versus total arrival rate is shown in Fig. \ref{fig:sims_one_queue_modelI}(a). We assume that $\lambda_1=\lambda_2=\lambda_3$.

%

It can be observed from Fig. \ref{fig:sims_one_queue_modelI}(a) that the average waiting time increases as the total arrival rate increases. This is because the congestion becomes worse as more vehicles enter the queue. Meanwhile, for a fixed total arrival rate, the average waiting time increases as the communication probability $p_t$ decreases, due to the increasing randomness of the traffic signal.

Fig. \ref{fig:sims_one_queue_modelI}(b) shows the average queue size versus communication probability $p_t$ over 10,000 time slots. It can be observed that the average queue size decreases as the communication probability $p_t$ increases. It is expected as when $p_t$ increases blocking probability reduces. This shows it is very important that vehicles communicate their intentions to the traffic light to avoid blocking.

\subsubsection{\modelII}
\begin{figure}[t!]
\vspace{10pt}
\begin{center}
\subfigure[Waiting time vs arrival rate]{ \scalebox{.30}{\includegraphics{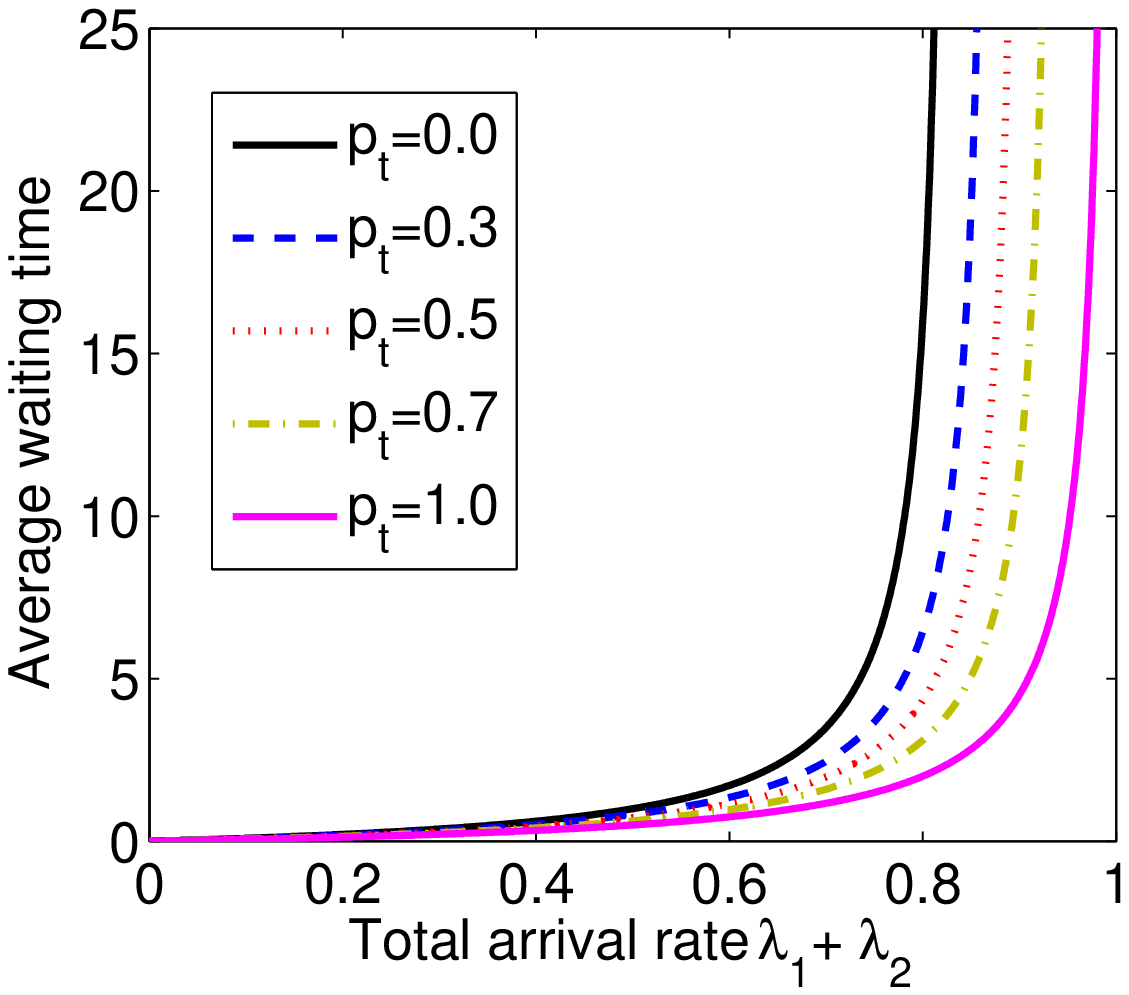}} } 
\subfigure[Queue Size vs $p_t$]{ \scalebox{.30}{\includegraphics{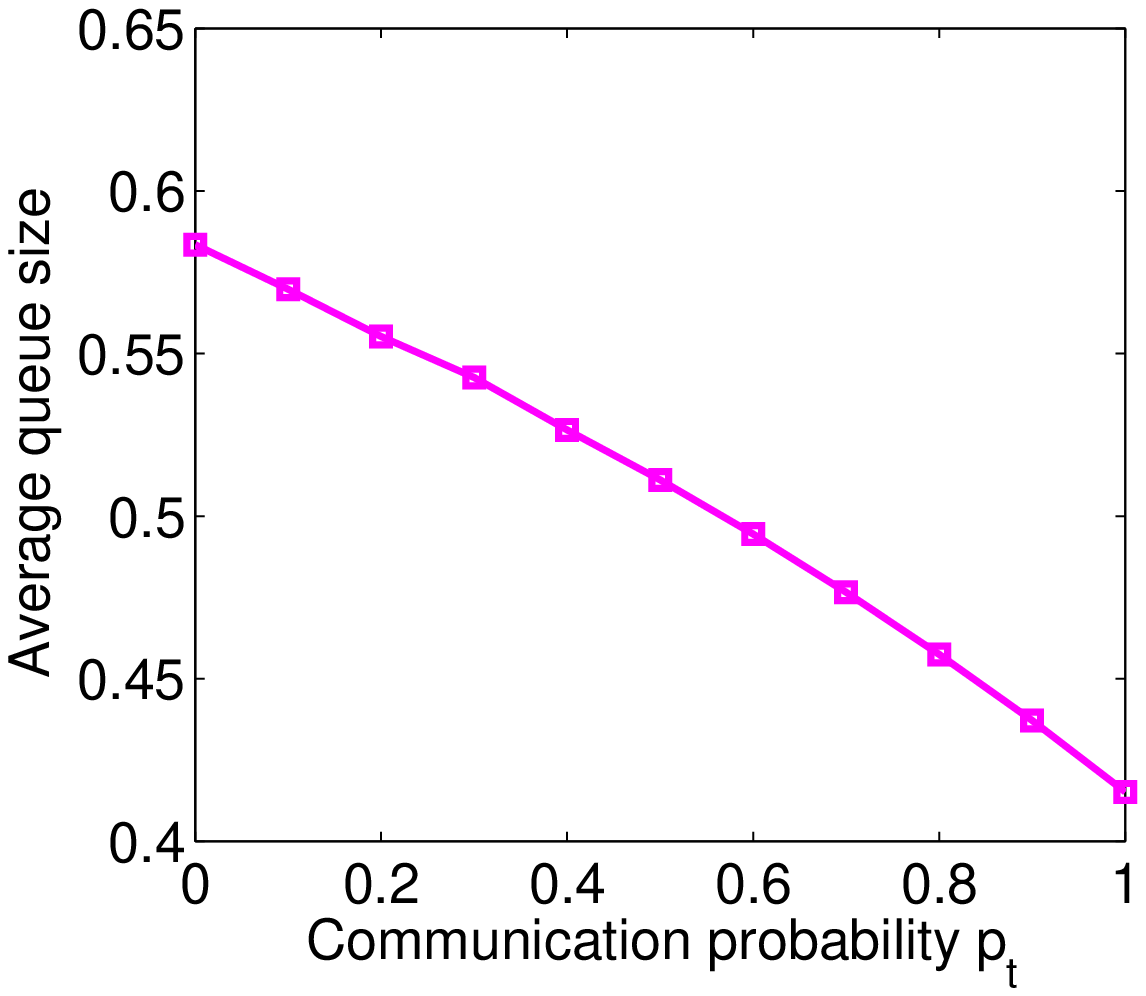}} } \\
\end{center}
\begin{center}
\vspace{-5pt}
\caption{\label{fig:sims_one_queue_modelII} \modelII. (a) Average waiting time for different total arrival rates with different communication probabilities $p_t$. (b) Average queue size versus communication probability $p_t$. The arrival rates are $\lambda_1=\lambda_2=0.15$.}
\vspace{-25pt}
\end{center}
\end{figure}

The simulated average waiting time for \modelII~ versus total arrival rate is shown in Fig. \ref{fig:sims_one_queue_modelII}(a), where we assume that $\lambda_1=\lambda_2$.

Fig. \ref{fig:sims_one_queue_modelII}(a) shows the similar relationship between total arrival rate and average waiting time as shown in Fig. \ref{fig:sims_one_queue_modelI}(a). However, it can be observed in Fig. \ref{fig:sims_one_queue_modelII}(a) that in most cases the average waiting time is much less than that in Fig. \ref{fig:sims_one_queue_modelI}(a) for the same total arrival rate and communication probability $p_t$ ($p_t\neq 1$). Only when $p_t=1$, the average waiting times are the same in Fig. \ref{fig:sims_one_queue_modelII}(a) and Fig. \ref{fig:sims_one_queue_modelI}(a) for the same total arrival rate. This is because when $p_t\neq 1$, at least one vehicle passes during two time slots in \modelII. However, it is possible that no vehicles can pass during several time slots in \modelI. Only when $p_t=1$, the service rate will always be 1 in both \modelI~ and \modelII, and their average waiting times are closer to each other. 

Fig. \ref{fig:sims_one_queue_modelII}(b) shows the average queue size versus communication probability $p_t$ over 10,000 time slots. In general, when communication probability $p_t$ is fixed, the average queue size in \modelII~ is smaller than that in \modelI. In addition, the average queue size in \modelII~ is almost linearly decreases as the communication probability $p_t$ increases.


\vspace{-5pt}
\subsection{Evaluation of the Shortest Delay Algorithm}
\begin{figure} [t!]
\vspace{10pt}
\centering
\scalebox{.41}{\includegraphics{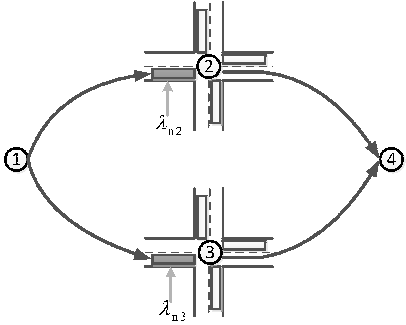}}
\vspace{-5pt}
\caption{An illustrative transportation network with four nodes. The total arrival rate to node 2 and node 3 is $\lambda_{n2}$ and $\lambda_{n3}$, respectively.}
\vspace{-5pt}
\label{roadmodel1}
\end{figure}

In this section, we evaluate our algorithm using an illustrative transportation network as shown in Fig. \ref{roadmodel1}.
In this network, there are four nodes, where node 1 and 4 are the starting and ending nodes, while node 2 and 3 are two intermediate intersections/nodes . The total arrival rate to node 2 and node 3 is $\lambda_{n2}$ and $\lambda_{n3}$ respectively. Assuming that a percentage of vehicles has communication abilities at node 2, and none of the vehicles have communication abilities at node 3, we evaluate the estimated end-to-end traveling time (or delay). Next, we briefly describe our baselines.

\subsubsection{Baselines}
We consider a baseline algorithm that uses the same method as our algorithm except that it does not take into account the communication probability of vehicles. In other words, none of the vehicles can communicate their intentions in the baseline algorithm. Thus, the evaluation of our shortest delay algorithm as compared to the baseline will show the benefit communication ability to overall delay, and avoiding blocking. The baseline algorithm is summarized briefly in the following.
\begin{itemize}
\item Given the arrival rate into each node, we calculate the average waiting time $W_n$ in each node without considering the communication ability of vehicles. For \modelI, $W_n$ is (\ref{waitingtime_model1_noonetells}) and for \modelII, $W_n$ is (\ref{waitingtime_model2_noonetells}).
\item Similar to our shortest delay algorithm, the total traveling time $T_m$ on road/edge $m$ is the sum of $W_n$ and $t_m$, \ie $T_m=W+t_m$, and $T_m $is assigned as the weight of edge $m$.
\item Given the starting and ending nodes, we run Dijkstra's algorithm and return the shortest path it finds.
\end{itemize}

Next, we present our simulation results for \modelI~ and  \modelII. In the transportation network in Fig.~\ref{roadmodel1}, we know that there are only two paths from node 1 to 4; they are $1\to 2\to 4$ and $1\to 3\to 4$ as shown in bold directed lines in Fig. \ref{roadmodel1}. To clearly see the effect of blocking and waiting times at intersections/queues, we assume that the four road segments in Fig. \ref{roadmodel1} have the same length. Thus, without violating generality, we assume that the traveling times over each road/link is 0, \ie both in the baseline and our algorithm $t_m=0$ ($m=1,2,3,4$).

\subsubsection{\modelI}
In this section, we consider that all the queuing model in the intersections of the road network follow \modelI, which is shown in Fig. \ref{fig:queueModels}(a).

\begin{figure}[t!]
\vspace{10pt}
\begin{center}
\subfigure[Traveling time vs $p_t$ at node 2]{ \scalebox{.30}{\includegraphics{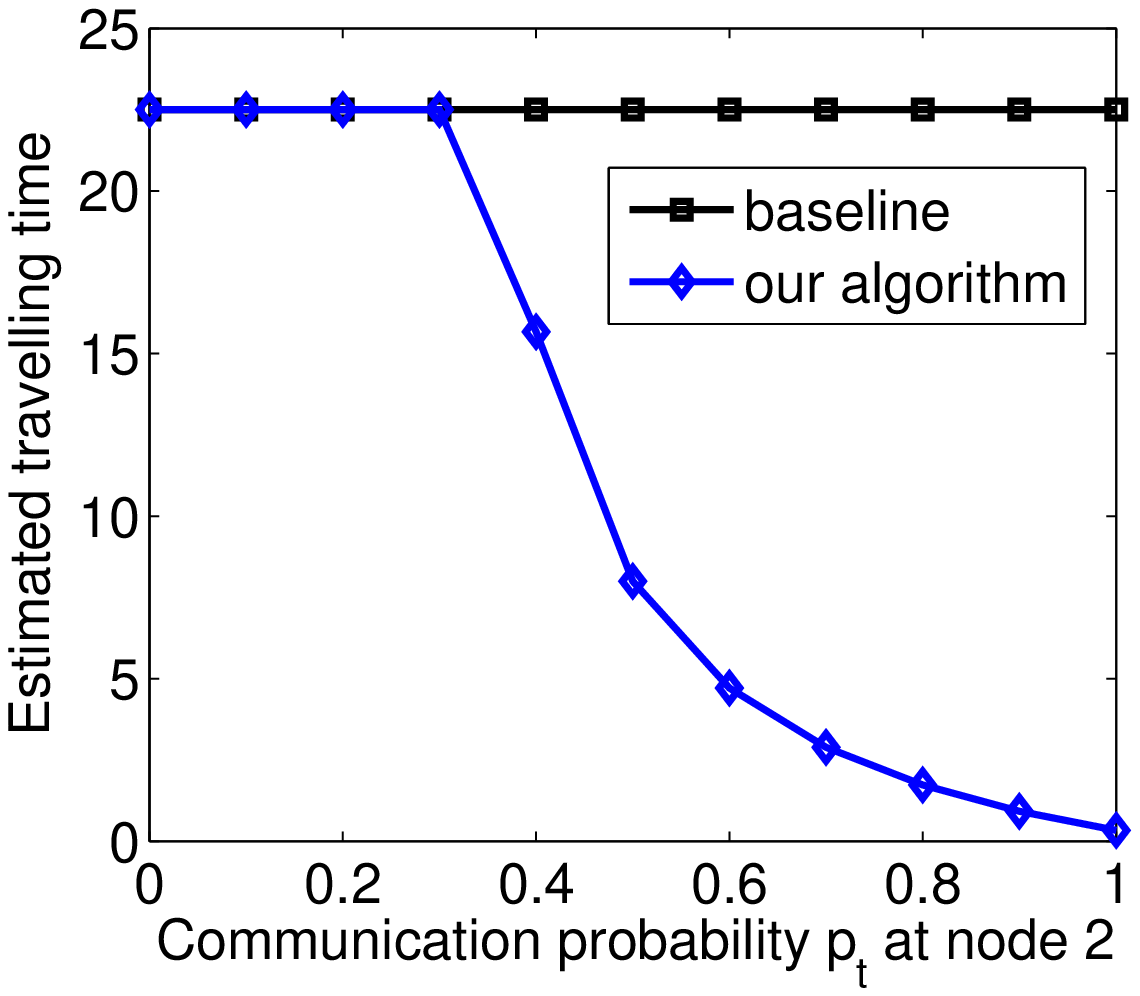}} } 
\subfigure[Traveling time vs $\lambda_{n2}$]{ \scalebox{.30}{\includegraphics{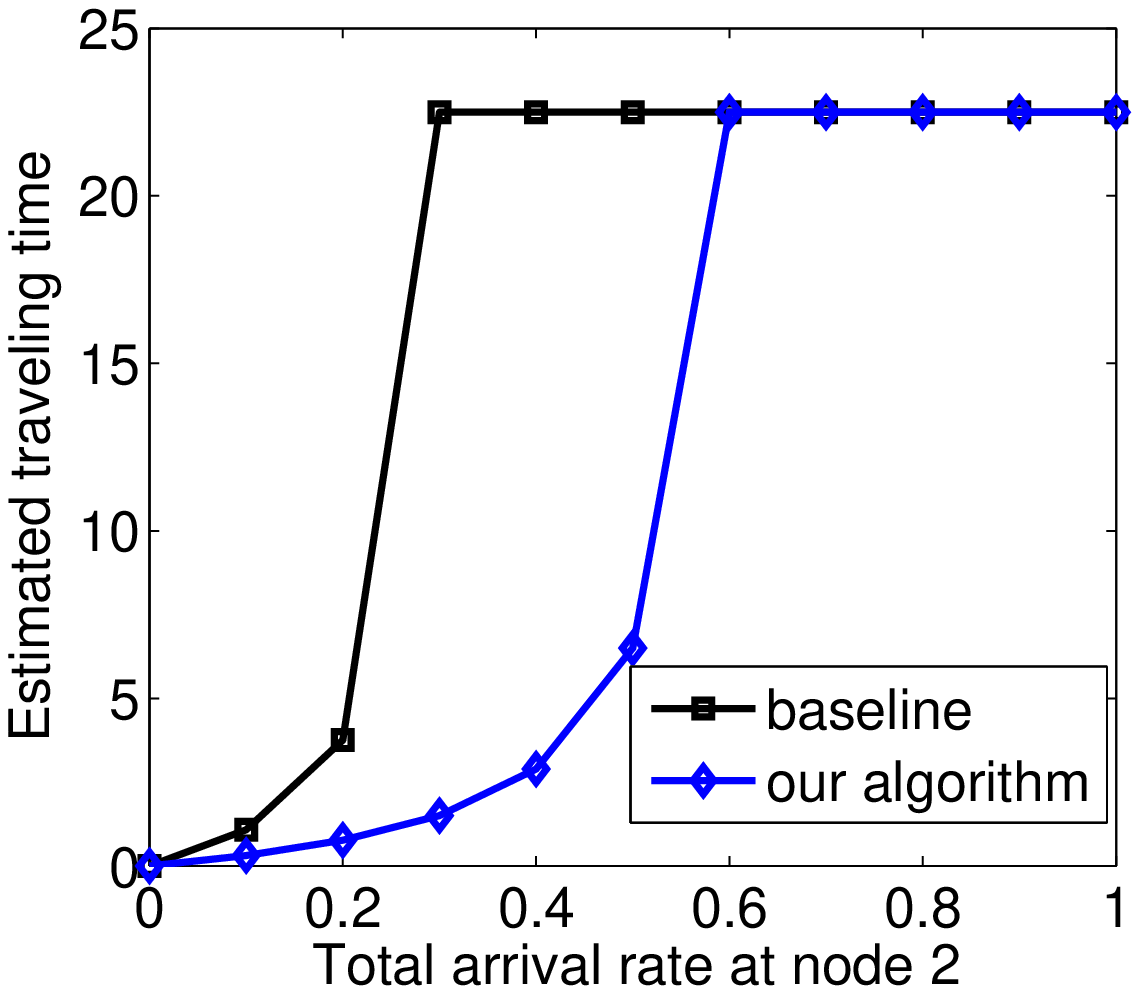}} } \\
\end{center}
\begin{center}
\vspace{-5pt}
\caption{\label{fig:sim_4nodes2paths_modelI} \modelI. (a) Estimated traveling time vs communication probability $p_t$ at node 2. The arrival rates at node 2 is $\lambda_1=\lambda_2=\lambda_3=0.4/3$, and the arrival rates at node 3 is $\lambda_1=\lambda_2=\lambda_3=0.1$. (b) Estimated traveling time vs arrival rate at node 2. The arrival rates at node 3 is always $\lambda_1=\lambda_2=\lambda_3=0.1$, and the communication probability $p_t$ at node 2 is 0.7.}
\vspace{-15pt}
\end{center}
\end{figure}


We assume that the arrival rates to node 2 are $\lambda_1=\lambda_2=\lambda_3=0.4/3$, and the arrival rates to node 3 are $\lambda_1=\lambda_2=\lambda_3=0.1$. Fig. \ref{fig:sim_4nodes2paths_modelI}(a) presents the traveling time versus communication probability $p_t$ at node 2. Since the baseline algorithm does not take into account the communication ability of vehicles, it will choose the path $1\to 3 \to 4$ as the shortest path since the total arrival rate at node 3 is smaller than node 2. However, if we take into account the communication ability of vehicles (\ie if we use our algorithm), Fig. \ref{fig:sim_4nodes2paths_modelI}(a) shows that the traveling time will decrease as our algorithm chooses the path $1\to 2\to 4$ instead of $1\to 3\to 4$ when the communication probability $p_t$ at node 2 is larger than 0.3.


Let us assume that the arrival rates to node 3 are $\lambda_1=\lambda_2=\lambda_3=0.1$, and the communication probability at node 2 is $p_t=0.7$. Fig. \ref{fig:sim_4nodes2paths_modelI}(b) presents the traveling time versus total arrival rates at node 2 (we assume $\lambda_1=\lambda_2=\lambda_3$ at node 2). Again, since the baseline algorithm does not take into account the communication ability of vehicles, it will choose the path $1\to 2 \to 4$ when $\lambda_{n2} \leq \lambda_{n3}$ and switches to path $1\to 3 \to 4$ when $\lambda_{n2} > \lambda_{n3}$. However, our algorithm chooses path $1\to 2 \to 4$ even for $\lambda_{n3}<\lambda_{n2}<0.6$ because of taking into account the communication ability of vehicles in node 2. Thus, the estimated traveling time using our algorithm is smaller than the baseline for $\lambda_{n2}<0.6$.

\subsubsection{\modelII}
In this section, we consider that all the queuing model in the intersections of the transportation network  follow \modelII, which is shown in Fig. \ref{model2}.

\begin{figure}[t!]
\begin{center}
\subfigure[Traveling time vs $p_t$ at node 2]{ \scalebox{.30}{\includegraphics{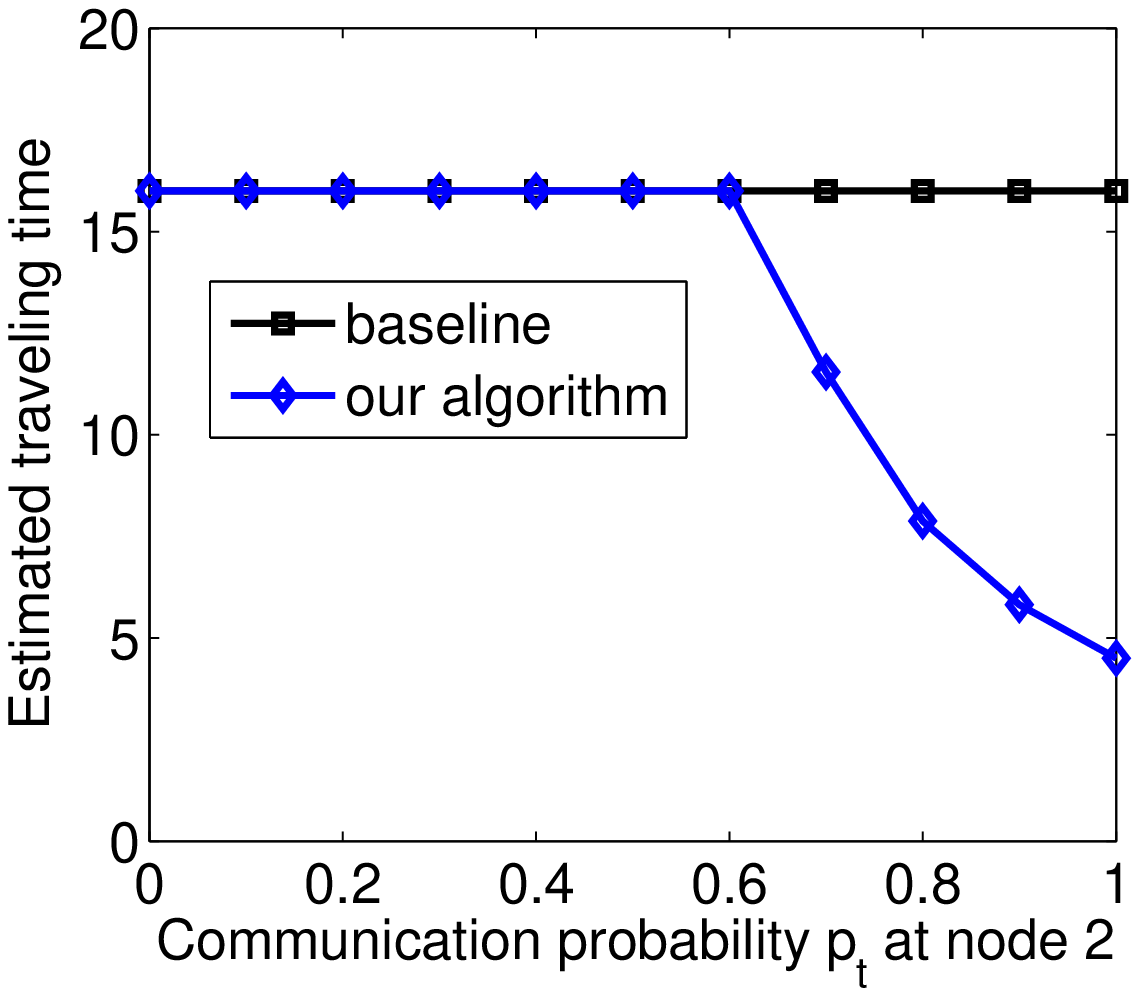}} } 
\subfigure[Traveling time vs $\lambda_{n2}$]{ \scalebox{.30}{\includegraphics{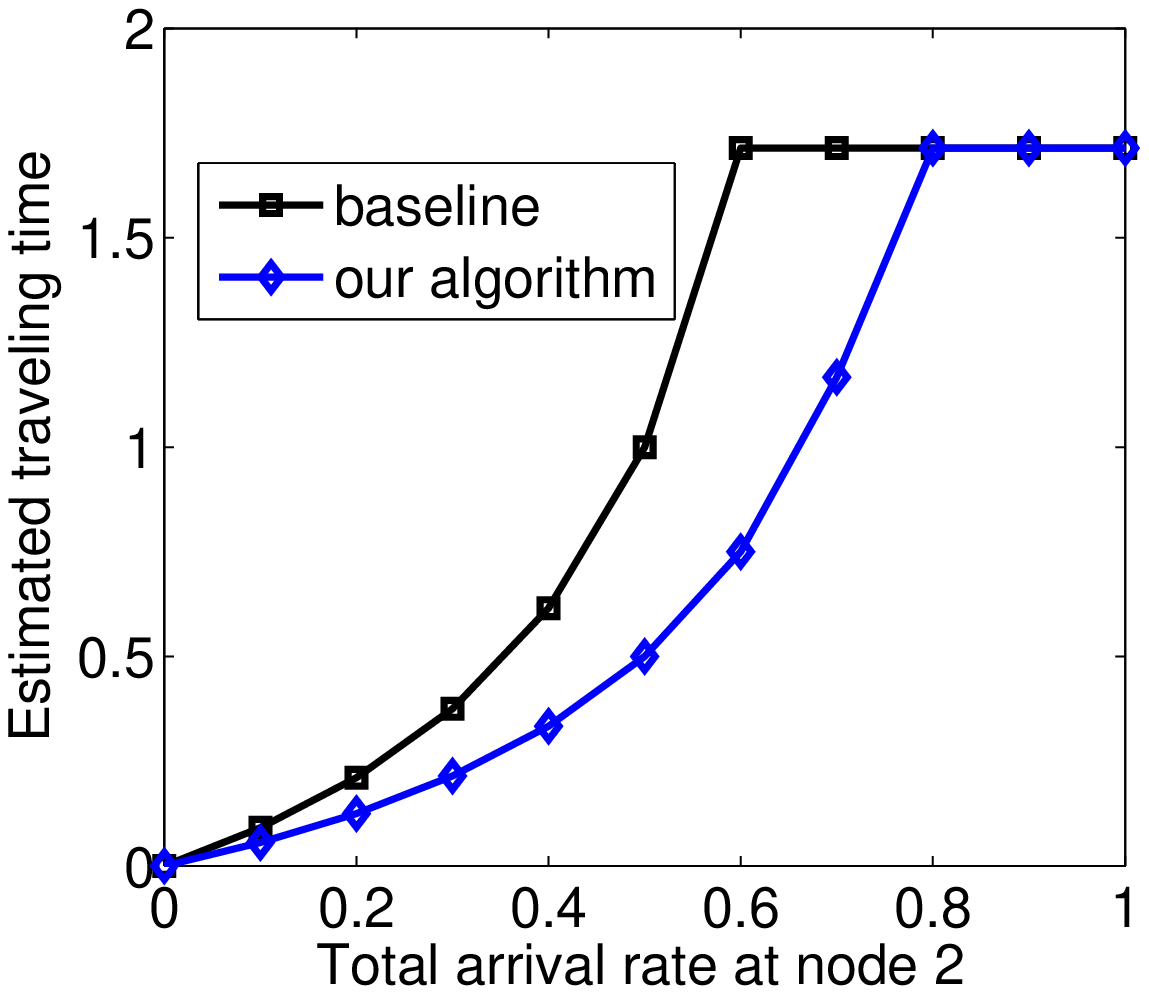}} } \\
\end{center}
\begin{center}
\vspace{-5pt}
\caption{\label{fig:sim_4nodes2paths_modelII} \modelII. (a) Estimated traveling time vs communication probability $p_t$ at node 2. The arrival rates at node 2 is $\lambda_1=\lambda_2=0.45$, and the arrival rates at node 3 is $\lambda_1=\lambda_2=0.4$. (b) Estimated traveling time vs arrival rate at node 2. The arrival rates at node 3 is always $\lambda_1=\lambda_2=0.3$, and the communication probability $p_t$ at node 2 is 1.0.}
\vspace{-25pt}
\end{center}
\end{figure}


Let us assume that the arrival rates to node 2 is $\lambda_1=\lambda_2=0.45$, and the arrival rates to node 3 is $\lambda_1=\lambda_2=0.4$. Fig. \ref{fig:sim_4nodes2paths_modelII}(a) presents the
traveling time versus communication probability $p_t$ at node 2. Similar to \modelI, our algorithm improves over the baseline.


Let us assume that the arrival rates to node 3 are $\lambda_1=\lambda_2=0.3$, and the communication probability at node 2 is $p_t=1.0$. Fig. \ref{fig:sim_4nodes2paths_modelII}(b) presents the traveling time versus total arrival rates at node 2 (we assume that $\lambda_1=\lambda_2$ at node 2). Similar to \modelI, our algorithm improves over the baseline. 

\vspace{-5pt}
\section{conclusion}\label{sec:conclusion}
In this paper, we investigated the blocking problem which naturally arises in transportation networks, where multiple vehicles with different itineraries share available resources. We characterized waiting times at intersections of transportation systems by taking into account blocking probability as well as the communication probability of vehicles. Then, by using average waiting times at intersection, we developed a shortest delay algorithm that calculates the routes with shortest delays between two points in a transportation network. Our simulation results show that our shortest delay algorithm significantly improves over baselines that are unaware of the blocking problem.

\vspace{-5pt}
\bibliographystyle{IEEEtran}

\end{document}